# An Automatic Validation System for Interferometry Density Measurements in the ENEA-FTU Tokamak Based on Soft- Computing[1]

G. Buceti, ENEA, Frascati, Italy
L. Fortuna, A. Rizzo, Università degli Studi di Catania, Catania, Italy
M.G. Xibilia, Università degli Studi di Messina, Messina, Italy

Abstract

In this paper, an automatic sensor validation strategy for the measurements of plasma line density in the ENEA-FTU tokamak is presented. Density measurements are performed by a 5-channel DCN interferometer. The approach proposed is based on the design of a neural model of the observed system., i.e. a model able to emulate the behavior of a fault-free sensor and of a two-stage fuzzy system able to detect the occurence of a fault by using a set of suitable indicators. The fault diagnosis and classification is also accomplished. The validation strategy has been implemented and embedded in an interactive software tool installed at FTU. Statistics concerning the rate of fault detection agree with the rate of uncertainty usually achieved in the post-pulse manual validation.

## 1 INTRODUCTION

With the continuous growth of the number of sensors installed in tokamaks and other big experimental physics plants, reliability of measurements has become a fundamental issue, both for feeding the control systems with reliable measurements and for physicians to analyse the real physics of the experiment. At present, most of the work is carried out manually by the experts responsible of the single diagnostic, which results in a very time-consuming activity. Moreover, knowledge about the validation process is spread among experts, each of them caring about peculiar aspects of the phenomena. Due to the nature of the knowledge, the huge amount of data stored and the heuristic involved, a suitable approach for the automatic validation of measurements has been individuated in the soft-computing-based techniques [1,2]. Some results have recently been achieved by the authors at JET, concerning the validation of the measurements of the vertical stresses on the bottom legs of the vacuum vessel during Vertical Displacement Events (VDEs) occurring at disruptions [3,4]. This papers deals with the design and implementation of an automatic validation tool for a 5-channel DCN interferometer installed at FTU-ENEA, Frascati.

The system is highly modular and organized in a hierarchical structure, providing advantages in terms of feasibility, flexibility of the validation actions, scalability and ease of upgrading.

## 2 THE VALIDATION SYSTEM

The first step in the design of the validation system consists of collecting information about the manual validation performed by experts. After thorough investigation, the following set of qualitative rules has been individuated:

1. line density is mainly related with the plasma current and with the quantity of gas injected into the tokamak;
2. line density is higher in the central area of the plasma column;
3. line density has a similar temporal trend on all channels (and similar to the plasma current trend, too);
4. line density trend is generally smooth: relevant discontinuities (step-like or spikes) are not admitted.

Based on this set of rules and on the wide set of experimental data available at FTU, the validation system reported in Fig. 1 has been designed. This is a modular system, consisting of two hierarchical stages.

The first stage consists of three main streams, each of them taking into account different aspects of validation coming from the rules illustrated above.

The first stream takes into account rules 1. and 2. To this aim, neural-network-based NARMAX models of each healthy DCN channel have been designed, based on a large set of manually validated data.

[1] Paper supported by MURST project 'Fault Detection and Diagnosis, Supervision and Control Reconfiguration in Industrial Process'

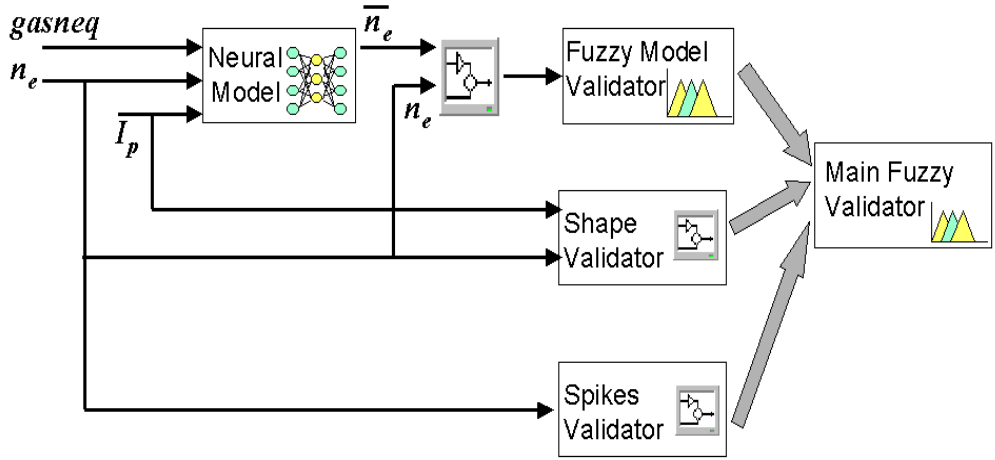

Figure 1: Block scheme of the sensor validation system.

Taking into account expert knowledge, through a wide series of network training phases, the following model has been considered for each channel:

$$d(k+1)=f(I_p(k+1), I_p(k), gas(k+1), gas(k), d(k), d(k-1))\quad(1)$$

where $k$ is the discrete time (sampling time 10 msec), $d(k)$ is the plasma density, $I_p(k)$ is the plasma current, $gas(k)$ is the total amount of gas introduced up to time $k$. The function $f$ has been approximated by a Multi-Layer-Perceptron [1], trained with a wide set of healty experimental data. Once the network has been trained, it can be used to detect the occurrence of faulty measurements by evaluating the residual between the ouput of the network and the actual measurement. As it is illustrated in Fig. 1, to increase the robustness of the residual evaluation, two further blocks are added. These are a simple signal processing block and the *Fuzzy Model Validator*. The signal processing block evaluates the average value of the residual (*avgerr*) and the maximum time interval in which the residual exceeds a certain threshold (*outbound*). The *Fuzzy Model Validator* is designed in order to provide a robust evaluation of the residual, by means of the following rule, in which the linguistic values are represented by suitable *fuzzy sets* [1].

R1: if *avgerr* is *zero* and *outbound* is *small* then *model* is *ok*

R2: if (*avgerr* is *positive* or *avgerr* is *negative*) and *outbound* is *small* then model is *warning*

R3: if (*avgerr* is *poshigh* or *avgerr* is *neghigh*) then *model* is *faulty*

R4: if (*avgerr* is *negative* or *avgerr* is *zero* or *avgerr* is *positive*) and *outbound* is *high* then *model* is *faulty*

R5: if (*avgerr* is *negative* or *avgerr* is zero or *avgerr* is *positive*) and *outbound* is *medium* then *model* is *warning*

The linguistic variable *model* indicates therefore whether the model output is in agreement with the sensor output, thus indicating that the sensor is working properly or not.

The second stream, *shape validator,* takes into account rule 3. The block evaluates the correlation between the trend of the plasma current and that of the line density coming from each channel of the interferometer and generates an indicator called *xc*.

The last stream, *spikes validator,* is designed to detect the occurrence of spikes or great discontinuities in the trend of measured line density. The block functionality is based on digital filters. The result of this validation is provided through an indicator called *spikes*.

Indicators coming from each of the three streams are fed as inputs to the second stage, *Main Fuzzy Validator*, in which a set of fuzzy rules combines all the partial validations performed by the previous stages. As a result, each channel is globally evaluated.

The set of fuzzy rules adopted for the global validation is the following:

R1: if *xc* is *high* and *model* is *ok* and *spikes* is *low* then *sensor* is *ok*

R2: if *xc* is *not-high* and *model* is *faulty* and *spikes* is *not-low* then *sensor* is *faulty*

R3: if *xc* is *low* and *model* is *ok* then *sensor* is *warning*

R4: if *xc* is *high* and *model* is *warning* and *spikes* is *low* then *sensor* is *ok*

R5: if *xc* is *medium* and *model* is *warning* then *sensor* is *warning*

R6: if *model* is *warning* and *spikes* is *not-low* then *sensor* is *faulty*

R7: if *xc* is *high* and *model* is *faulty* and *spikes* is *low* then *sensor* is *faulty*

R8: if *model* is *ok* and *spikes* is *not-low* then *sensor* is *warning*

R9: if *xc* is *low* and *model* is *ok* then *sensor* is *faulty*

R10: if *model* is *ok* and *spikes* is *very-high* then *sensor* is *faulty*

R11: if *xc* is *medium* and *model* is *ok* then *sensor* is *warning*

R12: if *model* is *ok* and *spikes* is *high* then *sensor* is *warning*

R13: if *xc* is *small* and *model* is *warning* then *sensor* is *faulty*

R14: if *xc* is *medium* and *model* is *faulty* and *spikes* is *not-low* then *sensor* is *faulty*

R15: if *xc* is *high* and *model* is *warning* and *spikes* is *medium* then *sensor* is *warning*

It can be noticed that a status of *warning* has been introduced, used to raise attention of the experts for a thorough analysis of the measurements of the shot.

Table 1 reports statistics on the performance of the validation tool, both in terms of *successful validation* (healthy measurements classified as healthy), and *successful fault detection* (faulty measurements classified as faulty), for each DCN channel, for a wide set of shots taken at different operational conditions.

Table 1: Margin specifications

| Channel # | Successful Validation | Successful Fault Detection |
|---|---|---|
| 1 | 92.4% | 95.5% |
| 2 | 90% | 95.2% |
| 3 | 94.3% | 72.2% |
| 4 | 90.9% | 98.1% |
| 5 | 85.3% | 85.9% |

## CONCLUSIONS

In this paper, a sensor validation tool for the DCN interferometer in the FTU tokamak in Frascati, Italy, has been presented. The tool is based on a cascade of a neural model and two fuzzy blocks. The neural model realises a NARMARX model of the plasma line density, taking as inputs the plasma current and the quantity of gas introduced into the tokamak. The output of the model is used to compute a set of suitable indicators which are used in a fuzzy block which gives a first validation of the sensor output. Subsequently, the information coming from the first fuzzy block is integrated in a second fuzzy validator to give a final validation of the sensor output. The designed validation tool has been tested on a large set of FTU data. The statistics on the overall performance have been reported in the paper for a large set of experiments, confirming the suitability of the approach.